# Scheduling Advertisement Delivery in Vehicular Networks

Gil Einziger, Carla-Fabiana Chiasserini, *Fellow, IEEE*, Francesco Malandrino, *Member, IEEE*

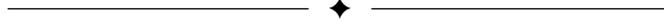

**Abstract**—Vehicular users are emerging as a prime market for targeted advertisement, where advertisements (ads) are sent from network points of access to vehicles, and displayed to passengers only if they are relevant to them. In this study, we take the viewpoint of a *broker* managing the advertisement system, and getting paid every time a relevant ad is displayed to an interested user. The broker selects the ads to broadcast at each point of access so as to maximize its revenue. In this context, we observe that choosing the ads that best fit the users' interest could actually hurt the broker's revenue. In light of this *conflict*, we present Volfied, an algorithm allowing for conflict-free, near-optimal ad selection with very low computational complexity. Our performance evaluation, carried out through real-world vehicular traces, shows that Volfied increases the broker revenue by up to 70% with provably low computational complexity, compared to state-of-the-art alternatives.

## 1 INTRODUCTION

Mobile devices have ceased to be a novelty: smartphones and tablets now dwarf TV sets and desktop computers as the preferred way to consume multimedia content [1]. Advertisers have followed suit, rapidly embracing a growing variety of mobile devices [2].

Among mobile devices, vehicles are playing an increasingly important role. The large amount of time drivers and passengers spend therein, coupled with the advanced capabilities of present-day and future on-board units, make them an ideal platform for mobile – especially location-specific – advertising. For instance, business such as shops and restaurants can advertise their products on the vehicles themselves, as an alternative to static billboards.

Modern advertisement, be it directed to website visitors or smartphone owners, is virtually always *targeted*. Popular services such as Google AdWords [3] and Ink TAD [4] build uncannily accurate [5] profiles of their users, with the purpose of only showing them *relevant* ads, matching their interests. The benefit of targeted advertisement is threefold [6]: advertisers make more potential customers aware of their products and services; brokers (e.g., Google) get higher click-through rates and thus higher profits; users are exposed to less disruptive, and potentially useful, content. Targeted advertisement is even more important in mobile scenarios, where screen sizes and attention span further limit the number of ads that users can visualize in a given time.

In our work, we consider a *vehicular advertisement* scenario. Similar to Internet advertising systems [7], [8], advertisers sign an agreement with a *broker*: upon successfully displaying a relevant ad to an interested user (called an *impression* [9]), the broker receives a payment determined by the *value* of the ad itself. In our vehicular environment, the broker controls a set of network points of access (PoAs), e.g., cellular base stations or DSRC roadside units, and decides which ads they should broadcast. Vehicles passing by a PoA get the ads, and display them to their drivers and/or passengers *only if* they are relevant to them. The number of impressions is then reported periodically (e.g., daily) through a secure connection.

In this context, the broker has to select in real time the ads to broadcast at each PoA, in order to maximize its own revenue. To this end, the broker can exploit information about the value of the ads and the users' interest (obtained through now-commonplace profiling techniques), as well as their trajectory and radio coverage (obtained from the PoAs themselves). At the same time, it has to account for the fact that (i) PoAs can only broadcast a limited number of ads, due to bandwidth considerations; (ii) vehicles will only display a limited number of ads to their drivers, so as to mitigate distractions, and (iii) such ads will be chosen by the vehicles based on how relevant they are to their drivers/passengers.

Our paper offers the following main contributions:

(i) We present a *system model* describing both the features of ads and the users' interests in a multi-dimensional space. Thanks to our model, we are able to identify a latent *conflict* existing between the users' interests and the broker's revenue, and show that strategies solely accounting for the former can severely hurt the latter. We therefore formulate an optimization problem, formalizing all aspects of the ad selection problem.

(ii) In view of the problem complexity, we present a *greedy algorithm*, named after the computer game *Volfied*.

(iii) We *formally prove* that (a) Volfied generates conflict-free decisions, (b) its performance never decreases as the number of ads to broadcast grows, and (c) it can be applied to a simplified system abstraction to obtain linear time complexity with a negligible performance loss.

(iv) We *evaluate* Volfied in a real-world scenario, comparing it against popular alternatives such as Top-k. We find that Volfied greatly outperforms the alternatives, increasing the broker's revenue by up to 70% and the number of impressions by almost 50%. Also, Volfied requires no *caching* at the vehicles, which reduces the hardware requirements imposed on vehicles, as well as the system implementation complexity. In particular, when Volfied is used, vehicles are not required to implement a caching policy or track the time and space

---

- All authors are with the Department of Electronics and Telecommunications at Politecnico di Torino, Torino, Italy. C. F. Chiasserini is also a Research Associate at IEIIT-CNR, Italy.

validity of each ad.

In spite of some common elements with previous works, as described in Sec. 2, our work presents a novel, *unified* model for vehicular advertising, capturing how network- and user-related constraints impact respectively the number of ads that brokers can transmit and users can visualize. Such a unified model is therefore able to identify the conflict between users' interests and broker's revenue. Similarly, the highly efficient ad selection algorithm we propose has, to the best of our knowledge, no parallel in the literature.

The remainder of the paper is organized as follows. After discussing related work in Sec. 2, we present our system model and optimization formulation in Sec. 3. Given the problem complexity, in Sec. 4 we present our online heuristics, Volfied, and highlight its properties. Then, in Sec. 5 we present a simplified system abstraction aimed at making the ad selection faster, and we quantify its impact on the broker's revenue. Finally, we present our reference scenario and our numerical results in Sec. 6, before concluding the paper in Sec. 7.

## 2 RELATED WORK

Our work is related to three main research areas: ad selection, mobile advertising, and content delivery in vehicular networks.

### 2.1 Ad selection and pricing

Advertisers and advertising platforms have to carefully choose *which* ads each user should be exposed to, in order to avoid irrelevant or overly annoying ads. This requires *a priori* [10] or *a posteriori* [11] estimation of how well ads will be received by the users they are displayed to. *A priori* estimation mostly uses social networks; the intuition behind [10] is that if a user's friends are interested in a certain product, then the user herself will likely be as well. [11] takes the a-posteriori approach, and aims at quantifying how well received a displayed ad was through such metrics as dwell time (how long users stay at the advertiser's site) and bounce rate (how likely they are to leave that site after only seeing one page). These issues are orthogonal to our scenario, where the user interests are part of the input.

A very important factor to account for when selecting ads is also *pricing*. Existing works aim at maximizing social welfare [12] in ad auctions, or take into account social influence in on-line advertising [8]. Another research direction is to treat selection of on-line ads as an optimization problem [9], [13], which however cannot scale to large systems due to solution complexity. Finally, an aspect that is out of the scope of this work, yet very relevant, is privacy in targeted advertising, which has been addressed in several papers (see, e.g., [7]).

### 2.2 Advertisement in mobile scenarios

Advertising to *mobile* users, e.g., drivers and passengers of a vehicle, poses several additional challenges, as the number of ads that can be delivered and displayed are limited by such factors as network capacity, storage space on mobile devices, and users' attention span. There is a body of works focusing on ad *scheduling*, i.e., deciding when (and where, in a mobile website or social network feed) ads should be displayed, and at which price. These works typically employ machine learning or game theory approaches [14], and either focus on ad pricing or attempt to maximize social welfare. For example, [15], [16] suggest methods for advertisers to configure their ads features and generation speed to maximize ads visibility. Similarly, [17] suggests ad scheduling techniques over a shared medium to maximize revenue. This approach requires advertisers to make complex decisions, which are not practical in a vehicular environment due to its real-time requirements.

Among the most recent works, [18] seeks to optimally choose the ads to interleave in a customized news feed. The problem is further decomposed into selecting which ads to serve (which is similar to the problem we address in our own study) and where to place the selected ads – an aspect which is not relevant in vehicular scenarios. Similarly, the authors of [19] study how to insert ads in a *streaming* scenario, e.g., a Facebook or Twitter newsfeed. While the problem they face shares some aspects with ours, namely, the on-line ad selection, the authors of [19] optimize a different metric (global welfare instead of broker's revenue) and use a different methodology (mechanism design instead of optimization) from our paper.

#### 2.2.1 Vehicular networks

Several recent works seek to leverage the unique conditions of vehicular networks, including mobility patterns and the availability of road-side infrastructure, to effectively serve ads to vehicular users. As an example, [20] uses mobility information to decide which (if any) ads any two vehicles should exchange upon meeting. The authors of [21] pursue a similar objective, and couple mobility information with signal-strength-based coding in order to cope with non-line-of-sight conditions. [22] takes one step further and looks into the *cause* of mobility, namely, the social ties between users: since users with similar interests are likely to move together, they can also exchange mutually-relevant ads. All of [20], [21], [22] use real-world datasets and mobility traces for their validation and performance evaluation.

Another body of works deals with *why* vehicular users should take part in the ad delivery process. In this context, [23] designs a cooperation framework where vehicles forward ads in exchange for virtual cash, which can be later redeemed for fuel or other services. Public-key cryptography, supported by road-side infrastructure, ensures that vehicles correctly report their behavior. Other works rely on game-theoretic approaches to foster cooperation; as an example, in [24] cooperative game theory is used to build a coalition of vehicles that cooperate with public buses in order to share ads.

Some works consider targeted advertisement, where ads are only relevant to certain vehicles. An example is [25], where ads are forwarded from road-side infrastructure to *seed* vehicles (selected based on their mobility), and from those to the other vehicles. The authors seek to maximize the number of displayed ads subject to a constraint on the number of seeds. Unlike our own work, [25] assumes that all relevant ads will eventually be displayed by vehicles, i.e., relevant ads are never discarded. The work in [26] addresses a scenario and problem similar to ours, but with a simpler display policy at the vehicle level. Their strategy is radically different from ours: decisions are made by solving an ILP optimization problem in real time, which would be infeasible in our scenario.

A preliminary version of our work has appeared in the conference paper [27], targeting the ad selection and delivery problem in a simpler setting and using a simpler system model. Major additions with respect to [27] include:

- a deeper discussion of the system architecture we consider, and its relationship with emerging paradigms like multi-access edge computing(Sec. 3.1), showing how Volfied can be applied in both present-day and future vehicular networks;
- a comparison of two similarity metrics, namely, the Euclidean distance and the cosine similarity (Sec. 3.2), highlighting how the performance of Volfied is not tied to a specific similarity metric;
- a formal proof that Volfied's performance never decreases as the number of ads to broadcast grows larger (Lemma 1 and Theorem 2 in Sec. 4.1.1), i.e., Volfied can make effective decisions in complex scenarios with multiple ads to choose from and each ad relevant to a different set of vehicles;
- a theoretical analysis of the impact of sparse-set approximation on performance (Sec. 5.2), bounding the revenue loss the Broker incurs, in exchange for lower complexity;
- an evaluation of the effect of caching on advertisement delivery (Sec. 6.3), showing that Volfied outperforms its counterparts even when vehicles are equipped with a cache.

## 2.3 Content delivery and infrastructure deployment

Mobile advertising is far from being the only use case for vehicular networks; there is a vast body of works studying how such networks can be leveraged to deliver, location-specific content of different nature [28], [29], [30], [31], [32], [33]. Earlier works focus on the viability of vehicular networks as a means for information storage and delivery [29] and on the comparison between different approaches to vehicle-assisted data delivery [30]. Later ones tend to propose more sophisticated schemes that use the available information on the network topology and vehicle mobility to either select the best-suited relays [31] or, using an opposite approach, move the content to an area where its recipient is likely to pass by [32]. Other works such as [33] aim at classifying existing proposals, and identifying the main strategies they adopt. Some studies take into account scenarios when there are *multiple* recipients for a specific content. In this case, it is important to select a *set* of relay nodes that will ease content dissemination [34], [35], without incurring in too much overhead. In a similar spirit, the authors of [36] investigate the concept of *vehicular content delivery network*.

*DSRC Road-side units* (RSUs) represent a unique feature of vehicular networks, and there are many works studying how they can be exploited for content caching [37] or cellular network offloading [38]. RSU *placement* is a fairly complex problem itself: the traditional approach is to make deployment decisions exploiting available information on vehicular mobility [39], [40] and/or possible vehicle-to-vehicle transfers. The problem is often mapped to well-known ones such as set covering or maximum coverage [35]. Recent works aim at optimizing application-specific quality-of-service metrics [41], [42] and present sophisticate solution techniques, from genetic algorithms [41] to polynomial-time, constant-rate approximation algorithms [43]. It is important to stress that the RSU deployment problem is orthogonal to ours; to us, the number and location of RSUs are input parameters.

## 3 SYSTEM SCENARIO AND PROBLEM FORMULATION

In this section, we first discuss our system architecture and how it integrates within present-day and future cellular networks (Sec. 3.1). Then we present the model of the system under study (Sec. 3.2) and define the problem of ad scheduling in vehicular networks (Sec. 3.3). Finally, in order to further clarify objective and system constraints, we provide a mathematical formulation of the problem whose solution provides the optimal ad selection, although at high complexity (Sec. 3.4).

### 3.1 System architecture

Our system architecture includes three main entities:
- vehicles, to which ads are pushed;
- PoAs, in charge of transmitting the ads to the vehicles;
- brokers, that controlling one or more PoAs.

It is the broker's job to select the sets of ads to be broadcasted at each time step by each PoA. In order to make this decision, the broker can use the following information:
- users' interests and preferences, stored in a *user information database*;
- the vehicles that are currently under coverage by PoAs and the PoAs visited by vehicles in the past;
- the ads that have been broadcasted by PoAs in the past, stored in an *ad registry*.

Information about users' interests can be provided by the users themselves upon subscription to the service, or obtained through nowadays-common profiling techniques. Information about the presence of vehicles under PoAs coverage can be obtained from the PoAs themselves, by exploiting the beacons vehicles periodically transmit [44]. Furthermore, brokers can have access to a *traffic server* containing information on the paths travelled by vehicles in the past, as standardized by 3GPP [45]. In our performance evaluation, we also study how errors in acquiring both user and traffic information affect the performance.

As far as the network architecture is concerned, both a centralized and a distributed scenario can be envisioned. Starting with the current LTE network architecture, depicted in Fig. 1(left), PoAs correspond to eNBs, which communicate with vehicles through LTE radio links. eNBs are then connected to the LTE network Evolved Packet Core (EPC), which contains such entities as the Serving Gateway (SGW) and the Packet Data Network Gateway (PDN-GW). Through the PDN-GW, data can reach the public Internet, including the datacenter (be it physical or virtual) hosting the broker and the auxiliary servers (traffic server, ad registry, and user information database). This architecture is centralized, due to the fact that no entities other than EPC network functions can be placed *within* the EPC network; in other words, data needs to traverse the PDN-GW before they can reach the broker or any other server. Similarly, we can consider DSRC RSUs that communicate with vehicles and are connected to the Internet as well, while the broker resides in a data center in the cloud.

An alternative to this scenario is the so-called Multi-access edge computing (MEC) architecture, which envisions endowing network edge nodes (e.g., eNBs or RSUs) with computation and storage capabilities, so as to process data close to the users (Fig. 1(right)). As an example, in this context eNBs can be integrated with a virtualized EPC (vEPC) instance and run a broker instance, implemented as a virtual application. Brokers controlling different eNBs can coordinate with each other, exchanging information about the ads they select. In this case, each broker would know which ads each other broker selects, thus making the ad server redundant.

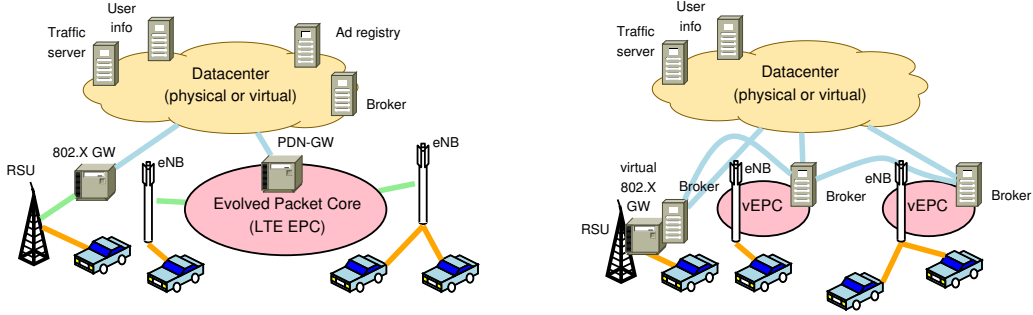

Fig. 1. System architecture including 802.11p and LTE radio links: centralized (left) and MEC (right). PoAs are either RSU or LTE eNBs. The broker resides in a datacenter in the cloud in the centralized scenario, while broker entities are located at the PoA in the MEC-based architecture. In the latter case, brokers can coordinate with each other and exchange information on the ads that the associated PoAs transmit. In this case, each broker would be informed on the ads transmitted by each other broker and the ad registry would be redundant; the communication with the datacenter would thus be limited to vehicle paths and user profiles.

### 3.2 System model

For simplicity, we consider that time is divided into discrete steps, whose duration should be chosen so as to match the features (e.g., vehicle mobility) of the scenario under consideration and, as we will see in Sec. 4, does not impact how our algorithms work or their complexity. At each time step, vehicles receive ads broadcasted by each PoA under whose coverage they pass, and can display up to $M$ ads that are relevant (i.e., of interest) to the user. Also, a vehicle displays an ad at most once[1]; the number of ads that can be broadcasted by PoAs is constrained by bandwidth and cost limitations. We denote the maximum number of ads that each PoA can broadcast in one time step by $K$.

**Model entities.** The main entities we need to model are: (i) *vehicles* (also referred as users), $v \in \mathcal{V}$, (ii) *ads* $a \in \mathcal{A}$ and (iii) PoAs $u \in \mathcal{U}$. Each ad $a$ has an associated *value*, $r(a, u)$. Having PoA-specific ad values allows us to model both *local* ads, which are worthless at PoAs out of their target location, and *global* ads, whose value is constant at all PoAs. Every time ad $a$ is displayed to a vehicle, the on-board platform notifies the broker, which gets a *revenue* equal to the ad value. Thus the broker's total revenue is given by the number of impressions, each weighted by the value of the displayed ad.

Vehicles can be equipped with a cache, allowing up to $C$ ads to be stored for later display. Unless otherwise specified, we assume that ads have to be displayed in the same time slot when they are downloaded ($C = 0$). The case where vehicles exploit a cache ($C > 0$), however, is considered in Sec. 6.

The content of ads and the interest of vehicles are both described in terms of *features*. Therefore, both ads and vehicles can be mapped onto points in an $n$-dimensional *feature space*, $\mathbb{F} \subseteq \mathbb{R}^n$, where $n$ is the number of features.

**Similarity and relevance.** Between any two features $\vec{f_1}, \vec{f_2}$ in the $\mathbb{F}$ space, we take a *distance* metric, $D(\vec{f_1}, \vec{f_2})$, as a measure of similarity. As it will be clearer later, this is instrumental in guaranteeing that conflicts between the users' interests and the broker's revenue are properly modeled and addressed.

In general, two points with similar locations in the $\mathbb{F}$ space will have a small distance (and we will also say that they are *close* to each other), while points with very different locations will have a large distance. The distance from a point to itself is always zero. In particular, in our system model, if both $\vec{f_1}$ and $\vec{f_2}$ are vehicles,

[1]. Ads that generate a revenue when displayed multiple times can be represented by separate elements of $\mathcal{A}$.

the distance expresses how different their interests are. If both are ads, $D(\vec{f_1}, \vec{f_2})$ conveys how different the ads themselves, and their potential audience, are. Finally, if $\vec{f_1}$ is an ad $a \in \mathcal{A}$ and $\vec{f_2}$ is a vehicle $v \in \mathcal{V}$, the distance $D(a, v)$ is related to how *relevant* ad $a$ is to vehicle $v$: the smaller $D(a, v)$ is, the more relevant $a$ is to $v$. We then define a *distance threshold* $D_{\max}$: only ads with a distance lower than $D_{\max}$ from user $v$, are relevant to $v$.

Note that multiple distance metrics have been proposed in the literature, aiming at capturing the different meanings of the concept of "closeness" between features in an $n$-dimensional space. It is important to point out that our system model and algorithm can work unmodified with any distance metric we choose, i.e., any metric that exhibits the triangle inequality property.

The most straightforward definition we can adopt for $D(\vec{f_1}, \vec{f_2})$ is the Euclidean distance:

$$D(\vec{f_1}, \vec{f_2}) = \left\| \vec{f_1} - \vec{f_2} \right\|_2. \qquad (1)$$

Cosine similarity [46] is often used to compare points in a multidimensional space, especially in machine learning applications. It is defined as:

$$\frac{\vec{f_1} \cdot \vec{f_2}}{\left\| \vec{f_1} \right\|_2 \left\| \vec{f_2} \right\|_2},$$

where $\left\| \cdot \right\|_2$ represents the 2-norm operator. The intuition behind the cosine similarity definition is to value orientation over magnitude. In our case, if two vehicles are interested in the same topics, they will have a high cosine similarity regardless of how strong each vehicle's interest is. Notice, however, that cosine similarity is not a distance metric. In applications where a distance metric is needed, *angular distance* is used instead [47]:

$$D(\vec{f_1}, \vec{f_2}) = \cos^{-1}\left(1 - \frac{\vec{f_1} \cdot \vec{f_2}}{\left\| \vec{f_1} \right\|_2 \left\| \vec{f_2} \right\|_2}\right), \qquad (2)$$

where $\cos^{-1}$ is the inverse cosine function. Note that, in accordance with our remark above, the larger the distance value in (2), the larger the difference between points $\vec{f_1}$ and $\vec{f_2}$.

The choice of the most appropriate distance metric depends on the individual scenario under consideration, and the semantics of the feature space $\mathcal{F}$. Cosine similarity and angular distance tend to work better where qualitative features (e.g., whether a given restaurant cooks Chinese food) dominate. On the other hand, quantitative aspects (e.g., prices or parking availability) tend to be best captured by the Euclidean distance. In our performance evaluation, we will use both the Euclidean distance (1) and the angular distance (2), checking that Volfied works properly regardless of the distance metric selected.

TABLE 1
Notation

| Symbol | Description |
|---|---|
| $\mathcal{A} = \{a\}$ | Set of advertisements |
| $\mathcal{V} = \{v\}$ | Set of vehicular users |
| $\mathcal{U} = \{u\}$ | Set of PoAs |
| $K$ | No. of ads each PoA can broadcast in one time step |
| $M$ | No. of ads each vehicle can display in one time step |
| $D(\cdot, \cdot)$ | Distance between users' interests and/or ads |
| $r(a, u)$ | Value of ad $a$ under the coverage of PoA $u$ |
| $R(a, u)$ | Estimated total revenue for ad $a \in \mathcal{A}$ at PoA $u$ |
| $\mathcal{A}_\epsilon^{(M)}$ | $M$-sparse approximation of set $\mathcal{A}$ |

The notation we use is summarized in Tab. 1.

### 3.3 Problem definition

As mentioned, it is the broker's task to define and enact what we formally call a *selection strategy*: given the set $\mathcal{A}$ of ads, the number of ads that can be broadcasted ($K$) and displayed ($M$), and the vehicles under PoA coverage, the broker has to select those ads that maximize its revenue. Intuitively, the broker should select ads that will be displayed by many vehicles and have a high value $r$. The former implies that the selected ads should be relevant *and* new to as many vehicles as possible, but, quite surprisingly, these two conditions are not sufficient to ensure that a broadcasted ad is actually displayed by the vehicles. Indeed, recall that the broker decides which ads to broadcast, but vehicles decide which of these ads to display. The aims of these two actors are different and potentially *conflicting*: the broker would aim at selecting ads with high value $r$, while vehicles display ads based on their relevance to the user. Thus, *whenever the broker can broadcast more ads than vehicles can display (i.e., $K > M$), a conflict between the broker and the users' interests may arise*. We remark that $K > M$ in all practical cases, and that, as highlighted in the example below, conflicts do not only waste radio resources, but they can also severely reduce the broker's revenue.

*Example 1 (Why conflicts hurt).* Consider a toy case with one PoA ($\mathcal{U} = \{u\}$), one vehicle ($\mathcal{V} = \{v\}$) and two ads ($\mathcal{A} = \{a_1, a_2\}$). Assume: $r(a_1) = 10$, $r(a_2) = 1$, $D(a_1, v) = 0.1$ and $D(a_2, v) = 0.05$. Also, let us focus on one time step and assume $M = 1$, i.e., the vehicle can display only one ad, and $D_{\max} = 0.15$. First, consider $K = 1$, i.e., the PoA can transmit only one ad, and that the PoA sends $a_1$. Then the vehicle will display $a_1$ and the broker will earn $r(a_1, u) = 10$. It is important to stress that the broker will select the ad to transmit in order to maximize its own revenue, and therefore it will send $a_1$ to $v$ *in spite* of the fact that $a_2$ would be more relevant. Now, assume $K = 2$ and that the PoA sends $a_1$ and $a_2$: one would expect that by sending more ads, the broker would earn *at least* the same revenue. However, owing to the fact that $M = 1$, vehicle $v$ will disregard $a_1$ and only display $a_2$, since $a_2$ is more relevant to $v$ than $a_1$. Thus, the broker's revenue will be $r(a_2, u) = 1$.

In light of this, we introduce the following definition.

*Definition 3.1 (conflict-free set).* A set of selected ads, $\mathcal{S} \subseteq \mathcal{A}$, is conflict free if, for each vehicle $v \in \mathcal{V}$, the set includes at most $M$ ads that are relevant to $v$.

### 3.4 Problem formulation

We now formally state the ad selection problem as follows. We denote the current time step by $t_c$, and the set of past and current steps by $\mathcal{T}$. Then the set of binary parameters $\chi(u, v, t) \in \{0, 1\}$ express whether PoA $u$ covers vehicle $v$ at time $t \in \mathcal{T}$.

Our formulation involves two binary decision variables: $\beta(a, u, t_c)$ and $\delta(a, v, t_c)$. The former concerns the broker, and it indicates whether an ad $a$ is broadcasted by PoA $u$ at the current time step or not. The latter concerns individual vehicles, and it indicates whether ad $a$ is displayed by vehicle $v$ at time $t_c$. Note that, although the vehicles and the broker make different decisions for different, and indeed conflicting, purposes, we are able to reproduce both decisions in the *same* optimization problem, as laid out next.

**Constraints.** A vehicle $v$ can display only the relevant ads that it receives from the current PoA, i.e., for any $a \in \mathcal{A}$ and $v \in \mathcal{V}$,

$$\delta(a, v, t_c) \leq \chi(u, v, t_c) \beta(a, u, t_c) \mathbb{1}_{[D(a,v) \leq D_{\max}]}, \quad (3)$$

where $u$ is the PoA, and $\mathbb{1}_{[D(a,v) \leq D_{\max}]}$ takes 1 if $D(a, v) \leq D_{\max}$ and 0 otherwise. Next, vehicles can display at most $M$ ads in time step $t_c$:

$$\sum_{a \in \mathcal{A}} \delta(a, v, t_c) \leq M, \quad \forall v \in \mathcal{V}. \quad (4)$$

Each ad can be shown at most once by every vehicle:

$$\sum_{t \in \mathcal{T}} \delta(a, v, t) \leq 1, \quad \forall a \in \mathcal{A}, v \in \mathcal{V}. \quad (5)$$

Note that the $\delta$ values that refer to previous time steps are input parameters to the problem.

Last, we must make sure that a vehicle $v$ selects the ads to display based on their relevance to itself. In other words, vehicle $v$ will not display an ad $a$ if it receives from the PoA $M$ (or more) ads whose relevance to $v$ is higher than $a$'s and were not displayed before:

$$\delta(a, v, t_c) \leq \max \left\{ 0, M - \sum_{\substack{a' \in \mathcal{A}: \\ D(a', v) > D(a, v)}} \left[ \chi(u, v, t_c) \beta(a', u, t_c) \left( 1 - \sum_{t \in \mathcal{T} \setminus \{t_c\}} \delta(a', v, t) \right) \right] \right\} \forall a \in \mathcal{A}, v \in \mathcal{V}. \quad (6)$$

Notice that the right-hand side of (6) can be greater than one; in that case, the constraint has no effect and the binary variable $\delta(a, v, t_c)$ can take value 0 or 1, whichever maximizes the objective.

As far as the broker is concerned, the only constraint is on the maximum number of ads that each PoA can broadcast at a given time step:

$$\sum_{a \in \mathcal{A}} \beta(a, u, t_c) \leq K \quad \forall u \in \mathcal{U}. \quad (7)$$

**Objective.** Given the above constraints, the broker's objective is to maximize its revenue at every time step:

$$\max \sum_{a \in \mathcal{A}} \sum_{v \in \mathcal{V}} \sum_{u \in \mathcal{U}} \delta(a, v, t_c) \chi(u, v, t_c) r(a, u). \quad (8)$$

**Discussion.** The above formulation has the interesting property of accounting for the way *both* vehicles and broker make decisions. Constraint (6) describes how vehicles will select ads based on the ads' relevance to themselves, while objective (8) represents the broker's aim to maximize its own revenue. Thus, conflicts are accounted for: by solving the optimization problem, the broker will maximize its revenue *subject to* the behavior of the vehicles.

On the negative side, the problem falls into the MILP (mixed-integer linear programming) category, and is therefore impractical to solve in real-time for moderate- to large-scale instances. Specifically, the ad selection is a $0-1$ knapsack problem with constant weights, whose item values are the outcome of another $0-1$ knapsack problem (the selection of the ads to display). Thus, the optimization problem is NP-hard. In light of this, we present below a heuristic approach.

## 4 ON-LINE DECISION MAKING

In this section, we present an *online* decision algorithm, allowing to make the same type of decisions as the optimization formulation introduced above, but (i) in a much more efficient way and (ii) without requiring any information on the future trajectories of vehicles. Clearly, while maximizing the objective in (8), subject to constraints (3)–(7) and given knowledge about all time steps in $\mathcal{T}$, ensures an *optimal* ad selection, our *online* decision algorithm may provide a suboptimal solution. We will show however that the online solution we obtain is optimal in some particular cases and closely matches the optimum under general conditions.

In order to design our online algorithm, we underline that the broker has to face two main challenges while selecting the ads to transmit. The first has to do with the *conflict* between the broker's revenue and the user interests, which may significantly impair the broker's revenue. The second is *complexity*, since the set of ads $\mathcal{A}$ is potentially very large, as are the sets of ads relevant to individual vehicles: this is especially important because we need to make our decisions using every time the most recent information available.

We address these two challenges separately. First, we propose a way to make *conflict-free* decisions leveraging on the estimated revenue that ads can generate. Then we introduce a *sparse-set* approximation that bounds the complexity of estimating ad revenues. For ease of presentation, we describe our decision-making scheme with reference to one PoA and one time step only, and we drop the PoA and time indices when discussing this scenario. Sec. 4.2 explains how to extend the proposed schemes to the multi-step and multi-PoA cases.

### 4.1 Conflict-free decisions: Volfied

In order to select a set of ads that maximizes its revenue, the broker has to first estimate the revenue it will get from broadcasting a generic ad $a$. Let $R(a)$ denote such estimated revenue. $R(a)$ is computed by adding $r(a)$ thereto every time a vehicle $v$, to which $a$ is relevant, enters the PoA coverage area, and subtracting the same amount when $v$ leaves the coverage area.

Armed with the estimated revenues $R(a)$, the broker applies an ad selection strategy. The most straightforward strategy would be *Top-k*, which selects the $K$ ads with highest estimated revenue $R(a)$. However, Top-k has the major disadvantage of ignoring the fact that vehicles can display at most $M$ ads each, thus it may create conflicts that harm the broker's revenue and waste radio resources on ads that will not be displayed (see Sec.

---

**Algorithm 1** Conflict-free ad selection: Volfied

**Require:** $\mathcal{A}, K, M, D_{\max}, R(a)$
1: $\mathcal{S} \leftarrow \emptyset$
2: **sort** $a \in \mathcal{A}$ **by** $R(a)$ in decreasing order
3: **for all** $a \in \mathcal{A}$ **do**
4:     **if** $|\{b \in \mathcal{S} \colon D(a,b) \leq 2D_{\max}\}| < M$ **then**
5:         $\mathcal{S} \leftarrow \mathcal{S} \cup \{a\}$
6:     **if** $|\mathcal{S}| \geq K$ **then**
7:         **break**
    **return** $\mathcal{S}$

---

3.3). To avoid this, we devise a *conflict-free* alternative, called Volfied and presented in Alg. 1.

The objective of Alg. 1 is to identify the set $\mathcal{S} \subseteq \mathcal{A}$ of ads to broadcast, initialized in line 1. Volfied starts by sorting set $\mathcal{A}$ by estimated revenue, in line 2. Then, for each ad $a$, it checks how many ads are already in $\mathcal{S}$ that are at a distance from $a$ smaller than $2D_{\max}$ (line 4). If less than $M$, $a$ is added to the set of ads to serve, in line 5. The algorithm ends when either all ads have been evaluated, or $K$ ads have been selected (line 6). By choosing a set of ads such that each has at most $M$ others within distance $2D_{\max}$ from itself, as we do in line 4–line 5, ensures us that there is no point in the whole space $\mathbb{F}$ that has more than $M$ ads within distance $D_{\max}$ from itself. Recalling that vehicles are also mapped to points in $\mathbb{F}$, this implies that no vehicle has more than $M$ relevant ads, i.e., the set chosen by Volfied is conflict-free.

In the following, we prove several important properties concerning the quality of the decisions made by Volfied (Sec. 4.1.1) and its time complexity (Sec. 4.1.2).

### 4.1.1 Quality of decisions

The most important property of the decisions made by Volfied is that they are conflict-free. Formally, the following theorem holds.

**Theorem 1.** *The set of ads $\mathcal{S}$ selected by Volfied is conflict free.*

*Proof:* Consider a set with one ad only; this is clearly conflict free. Then, by construction (line 4), Volfied selects an additional ad only if, for every ad $a \in \mathcal{S}$, there are less than $M$ ads within distance $2D_{\max}$. This implies that, for any vehicle $v$, $\mathcal{S}$ includes at most $M$ ads relevant to $v$, i.e., $\mathcal{S}$ remains conflict free. Indeed, due to triangle inequality, for any two ads $a$ and $b$ s.t. $D(a,b) > 2D_{\max}$, we have: $D(a,v) + D(v,b) \geq D(a,b) > 2D_{\max}$, for any vehicle $v$. That is, given an ad $a$, which is relevant to $v$, only ads within distance $2D_{\max}$ from $a$ may be relevant to $v$ too. □

We can also prove a *robustness* property, namely, that the performance of Volfied never decreases if we increase the number $K$ of ads to broadcast, i.e., it is *monotonic* with respect to it. To that end, we first need to prove the following lemma on the validity of revenue estimations.

**Lemma 1.** *The revenue obtained by broadcasting by a conflict-free set of ads $\mathcal{S} \subseteq \mathcal{A}$, with $|\mathcal{S}| \leq K$, corresponds to the sum of the estimated revenues of the ads therein, i.e., $\sum_{a \in \mathcal{S}} R(a)$.*

*Proof:* As discussed in Sec. 4.1, the estimated revenue $R(a)$ of each ad $a$ is the product between the value $r(a)$ and the number of vehicles for which $a$ is relevant. If the set $\mathcal{S}$ is conflict-free, then all vehicles will display all ads relevant to them, with each ad contributing exactly $R(a)$ to the broker's revenue. □

We can now prove that increasing the number $K$ of ads to select never hurts. To this end, it is sufficient to prove that increasing $K$ by one does not decrease the revenue.

**Theorem 2.** Increasing $K$ by one never hurts the total revenue yielded by Volfied.

*Proof:* Having a higher value of $K$ means that Volfied will run for more iterations. Since the set of ads and their revenue are unmodified, Volfied will still select the same $K_{\text{orig}}$ ads that it would select with a lower value of $K$, and (at most) one additional ad, i.e., $\mathcal{S}_{\text{orig}} \supseteq \mathcal{S}_{\text{new}}$. By Lemma 1, this also implies that the revenue does not decrease. □

It is important to notice that Theorem 2 depends on Lemma 1, i.e., on the fact that the sets of selected ads are conflict-free. Indeed, Example 1 depicts of how increasing $K$ can decrease the revenue if non-conflict-free decisions are made.

Finally, we show that, in the special case when $K = M$, Volfied is optimal:

**Property 1.** When $K = M$, then (i) Volfied and Top-k select the same set of ads, and (ii) such a set is optimal.

*Proof:* Both algorithms select the $\min\{K, |\mathcal{A}|\}$ highest-revenue ads. It follows that the condition in line 4 in Alg. 1 is always met (as $K = M$), thus Volfied and Top-k select the same ads. By Theorem 1, the set is conflict free; also it maximizes the broker's revenue since, by construction, it includes the $K$ ads with top estimated revenue $R(a)$. □

### 4.1.2 Complexity

We now turn to the issue of the time complexity of Volfied, and prove that it is linear in the size of the ads set.

**Theorem 3.** The worst-case runtime complexity of Volfied is $O(|\mathcal{A}| \cdot K)$.

*Proof:* From Algorithm 1, one can see that the loop in line 3 iterates over all the ads $a \in \mathcal{A}$, thus in the worst case all ads in $\mathcal{A}$ are processed. In line 4, we compare each ad against all previously selected ads, which are at most $K - 1$. The operations in the remaining lines have complexity $O(1)$ and thus the overall complexity of Algorithm 1 is $O(|\mathcal{A}| \cdot K)$. □

Finally, we remark that Volfied relies on the estimated revenues, i.e., the $R(a)$ values. Such estimates need to be refreshed every time a vehicle enters or exits the coverage area of a PoA. Every update has a linear cost in the number of ads, as shown below.

**Property 2.** The worst-case complexity of updating the revenue estimation, due to a vehicle entering/exiting a PoA coverage area, is $|\mathcal{A}|$.

*Proof:* Consider a vehicle $v$ and that all ads are relevant to $v$. When $v$ enters or leaves the coverage of a PoA, the revenue estimation of all ads (in the worst case) has to be updated, thus the complexity is $|\mathcal{A}|$. □

The *overall* complexity of Volfied coupled with the revenue estimation procedure is then $O(K \cdot |\mathcal{A}|^2)$. Clearly, when the number of ads and vehicles involved is large, the update procedure becomes cumbersome. To overcome this issue, below we propose an efficient approach which greatly reduces the number of ads to consider, as described in Sec. 5.

It is worth mentioning that the duration of time steps impacts none of the complexity results stated above. Shorter time steps will indeed imply that Volfied is run more frequently, but the complexity of each run remains very low, as stated in Theorem 3. As far as the revenue estimation procedure is concerned, it is run every time a vehicle enters or exists the coverage area of a PoA, and therefore it is not affected by the time step duration.

### 4.2 Multi-PoA, multi-step

The ad selection algorithm can be easily extended to networks comprising multiple PoAs and operating for multiple time steps, such as the one considered in our performance evaluation in Sec. 6.

Specifically, when considering multiple time steps, there is no profit in serving vehicles with the same ad multiple times. Thus, the way the estimated revenues $R$ are computed is enhanced as follows: $R(a, u)$ is increased by $r(a, u)$ only if $a$ has not been broadcasted to the vehicle before.

Similarly, we can account for the presence of multiple PoAs, i.e., for the fact that vehicles may have received an ad from some PoA they visited in the past. If a vehicle under the coverage of a PoA, $u \in \mathcal{U}$, has been served ad $a \in \mathcal{A}$ in the past by another PoA, $u' \in \mathcal{U}$, the corresponding $r(a, u)$ value is discounted from the revenue estimates $R(a, u)$. This requires the broker (not the advertisers) to know which PoAs the vehicles visited, a piece of information that can be easily gathered from the beacons cars are required to periodically send and that will be available in next-generation network systems [44], or by a *traffic server* entity as specified by 3GPP [45]. It does *not* require to know which ads were displayed by cars.

Notice that the Volfied algorithm itself works unmodified in multi-PoA and multi-step scenarios. In other words, the *algorithm* always makes per-PoA decisions, whether the *architecture* is centralized (as in Fig. 1(left)) or decentralized (as in Fig. 1(right)). It also implies that all the properties we proved above still hold. Nevertheless, the PoA deployment does impact the performance of vehicular advertising, as we show in Sec. 6.3.

## 5 REDUCING THE NUMBER OF ADS

On-line approaches such as Volfied require making decisions as they are needed, using every time the most recent available information – in our case, the revenue estimates $R(a)$. These estimates must be refreshed every time a vehicle enters or exits the coverage area of a PoA and, as stated in Property 2, the complexity of doing so is linear in the number $|\mathcal{A}|$ of ads. Such a complexity can be too high in some scenarios with many ads and/or high vehicular mobility.

To cope with this issue, in Sec. 5.1 we present a technique aimed at replacing the original set of ads with a *sparse* approximation thereof, where the number of ads relevant to any given vehicle is bounded. Then, in Sec. 5.2, we formally prove a bound to the revenue loss we can incur into by using the sparse ad set *in lieu* of the original one.

### 5.1 Computing the sparse-set approximation

As mentioned earlier, the complexity of the revenue estimation procedure grows linearly with the size of the set $\mathcal{A}$ of ads. A straightforward solution to speed up the ad selection procedure consists in limiting the size of such set. However, blindly removing ads would wantonly impair the system performance: the problem is not that there are too many ads, but there are too many ads *similar to each other*, hence with the same target audience. We therefore replace the set $\mathcal{A}$ with its *sparse approximation*, as defined below. For the sake of clarity, we start by considering $M = 1$, i.e., each vehicle can display at most one ad per time step.

***Definition 5.1 (Sparse set).*** $\mathcal{X} \subseteq \mathbb{F}$ is a sparse set if, for any two points $\vec{f_1}, \vec{f_2} \in \mathcal{X}$, $D(\vec{f_1}, \vec{f_2}) > 2\epsilon$.

The following result states that, given a sparse set of ads $\mathcal{X} \subseteq \mathbb{F}$, the distance between a point in $\mathcal{X}$ and any other point either in $\mathcal{X}$ or in $\mathcal{V}$ ($\mathcal{V} \subseteq \mathbb{F}$), is at least $\epsilon$. It follows that, given $D_{\max}$, a vehicle cannot find in $\mathcal{X}$ more than $\lceil (D_{\max}/\epsilon)^n \rceil$ ads that are relevant to itself.

**Theorem 4.** Given a sparse set $\mathcal{X} \subseteq \mathcal{A}$, for every point $\vec{f} \in \mathcal{X} \cup \mathcal{V}$, a closed ball of radius $\epsilon$ around $\vec{f}$ contains at most a single ad $a \in \mathcal{X}$.

*Proof:* In the case where $\vec{f} \in \mathcal{X}$ is an ad, the theorem holds given the definition of sparse set. Next, consider that $\vec{f}$ is a vehicle; in this case the proof is given by contradiction, as follows. Assume that there are two ads $a_1$ and $a_2$ in $\mathcal{X}$ s.t. $D(a_1, \vec{f}) \leq \epsilon$ and $D(a_2, \vec{f}) \leq \epsilon$. Then, by triangular inequality, $D(a_1, a_2) \leq D(a_1, \vec{f}) + D(a_2, \vec{f}) \leq 2\epsilon$, which contradicts the definition of sparse set. Thus the thesis is proven. □

Let us now introduce the sparse approximation of an ad set.

***Definition 5.2 (Sparse approximation).*** The sparse approximation of a set of ads $\mathcal{A}$ is a set $\mathcal{A}_\epsilon^{(1)} \subseteq \mathcal{A}$ such that: (i) $\mathcal{A}_\epsilon^{(1)}$ is a sparse set, and (ii) for each ad $a \in \mathcal{A}$, there exists $a' \in \mathcal{A}_\epsilon^{(1)}$ with $r(a') \geq r(a)$ and $D(a, a') \leq 2\epsilon$.

It is important to note that, due to local ads that are relevant only to vehicles under PoAs located within the ads target location, different PoAs may select different ads to be part of their sparse approximation.

Intuitively, $\mathcal{A}_\epsilon^{(1)}$ is a sparse set obtained by removing redundant, low-value ads from $\mathcal{A}$. Alg. 2 provides a technique to build the sparse approximation of $\mathcal{A}$. The algorithm first sorts

**Algorithm 2** Building a sparse approximation of the ad set (function EpsilonSet)

**Require:** $\mathcal{A}, \epsilon$
1:  $\mathcal{A}_\epsilon^{(1)} \leftarrow \emptyset$
2:  **sort** $a \in \mathcal{A}$ **by** $r(a)$ in decreasing order
3:  **for all** $a \in \mathcal{A}$ **do**
4:       $\mathcal{A}_\epsilon^{(1)} \leftarrow \mathcal{A}_\epsilon^{(1)} \cup \{a\}$
5:       $\mathcal{A} \leftarrow \mathcal{A} \setminus \{b \in \mathcal{A} : D(a,b) \leq 2\epsilon\}$
     **return** $\mathcal{A}_\epsilon^{(1)}$

the ads in the original set by their value (line 2). Then, at each iteration, it adds the top (i.e., highest-value) ad to the sparse set (line 4) and removes all other ads in $\mathcal{A}$ at distance less than or equal to $2\epsilon$ from said ad (line 5). An example of how Alg. 2 works is presented in Fig. 2. It is straightforward to see that, by construction, the resulting set $\mathcal{A}_\epsilon^{(1)}$ is the sparse approximation of $\mathcal{A}$, as by Definition 5.2.

Next, we consider $M > 1$ (i.e., vehicles can display more than one ad per time step). In this case, the broker should select multiple ads targeting the same audience, if the value of $K$ allows it. We therefore introduce the notion of $M$-sparse set and $M$-sparse approximation, $\mathcal{A}_\epsilon^{(M)}$.

***Definition 5.3 ($M$-sparse set).*** $\mathcal{X}^{(M)} \subseteq \mathbb{F}$ is an $M$-sparse set if, for any point $\vec{f} \in \mathcal{X}^{(M)}$, there are at most $M$ points within distance $2\epsilon$ from $\vec{f}$ (including $\vec{f}$ itself).

***Definition 5.4 ($M$-sparse approximation).*** The $M$-sparse approximation of a set of ads $\mathcal{A}$ is a set $\mathcal{A}_\epsilon^{(M)} \subseteq \mathcal{A}$ such that (i) $\mathcal{A}_\epsilon^{(M)}$ is $M$-sparse, and (ii) for each subset $\mathcal{B} \subseteq \mathcal{A}$ with $|\mathcal{B}| \leq M$, there exists a subset $\mathcal{B}_\epsilon^{(M)} \subseteq \mathcal{A}_\epsilon^{(M)}$ with $|\mathcal{B}_\epsilon^{(M)}| = |\mathcal{B}|$ and bijection function, $g : \mathcal{B} \to \mathcal{B}_\epsilon^{(M)}$, s.t. $\forall b \in \mathcal{B}: r(g(b)) \geq r(b)$ and $D(b, g(b)) \leq 2\epsilon$.

**Algorithm 3** Building the $M$-sparse approximation of the ad set

**Require:** $\mathcal{A}, \epsilon, M$
1:  $\mathcal{A}_\epsilon^{(0)} \leftarrow \emptyset$
2:  **for** $j = 1$ **to** $M$ **do**
3:       $\mathcal{A}_\epsilon^{(j)} \leftarrow \mathcal{A}_\epsilon^{(j-1)} \cup \texttt{EpsilonSet}(\mathcal{A} \setminus \mathcal{A}_\epsilon^{(j-1)}, \epsilon)$
     **return** $\mathcal{A}_\epsilon^{(M)}$

It is easy to see that, by construction, Alg. 3 builds the $M$-sparse approximation of the ad set $\mathcal{A}$. Indeed, it repeatedly calls the EpsilonSet function defined in Alg. 2. As shown by the following theorem, the sparse set resulting from Alg. 3 includes groups of up to $M$ similar ads that are relevant to the same vehicle.

**Theorem 5.** Given an $M$-sparse set $\mathcal{A}_\varepsilon^{(M)}$ output by Alg. 3, for every vehicle $v \in \mathcal{V}$, a closed ball of radius $\epsilon$ around $v$ contains at most $M$ ads.

*Proof:* The set $\mathcal{A}_\varepsilon^{(M)}$ is generated recursively by forming $M$ sparse sets (as it can be seen in Algorithm 3). Each sparse set satisfies Theorem 4 and, thus, contributes with at most a single ad s.t. $D(a, v) \leq \epsilon$. It follows that the maximum number of ads within a closed ball of radius $\epsilon$, centered in $v$, is equal to $M$. □

Replacing the original set of ads $\mathcal{A}$ with its sparse approximation $\mathcal{A}_\epsilon^{(M)}$ makes it possible for the broker to streamline the ad selection procedure. In particular, the estimate of the revenue, $R(a)$, can be updated with bounded complexity.

**Theorem 6.** When performed on $\mathcal{A}_\epsilon^{(M)}$, the complexity of the revenue estimation update is: $\min \left\{ \left\lceil \left( \frac{M \cdot D_{\max}}{\epsilon} \right)^n \right\rceil, \left| \mathcal{A}_\epsilon^{(M)} \right| \right\}$.

*Proof:* It follows from Theorem 5 that for each vehicle $v$ a closed ball of radius $\epsilon$ around $v$ contains at most $M$ ads. Therefore, we are left to consider how many such balls fit into a closed ball of radius $D_{\max}$. The maximum number of balls of radius $\epsilon$ that fit in such a volume is: $\left\lceil \left( \frac{M \cdot D_{\max}}{\epsilon} \right)^n \right\rceil$. Thus the maximum number of ads within distance $D_{\max}$ from the vehicle is the minimum between such a value and the total number of ads in $\mathcal{A}_\epsilon^{(M)}$. □

Clearly, larger values of $\epsilon$ allow a greater reduction of the number of ads, hence a faster ad selection. However, as $\epsilon$ grows, $a \in \mathcal{A}$ and its corresponding ad, $a' \in \mathcal{A}_\epsilon^{(M)}$, become less similar. It follows that $a'$ may become not relevant to a certain vehicle (i.e., $D(a', v) > D_{\max}$) while $a$ was (i.e., $D(a, v) \leq D_{\max}$). This means that the opportunities of a selection strategy to pick $M$ relevant ads for a vehicle may diminish when the strategy is applied to $\mathcal{A}_\epsilon^{(M)}$ instead of $\mathcal{A}$. This may lead to a revenue loss, which is analyzed in detail in Sec. 5.2. Also, in Sec. 6 we show that such a performance loss is negligible even for large values of $\epsilon$, e.g., $\epsilon = D_{\max}/4$.

Finally, we remark that the sparse approximation of the ad set needs to be computed only *once* (and can be computed offline), while the selection algorithm runs every time a new set of ads to be broadcasted has to be identified.

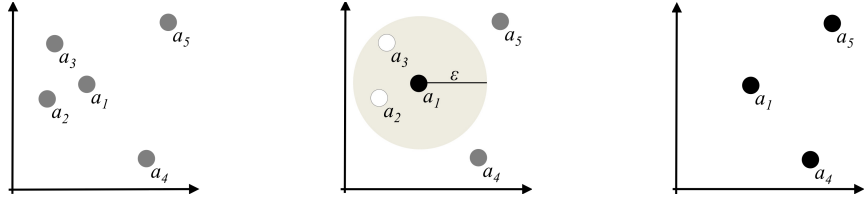

Fig. 2. How Alg. 2 works: given the initial set $\mathcal{A}$ with $r(a_1) > r(a_2) > \ldots > r(a_5)$ (left), it first includes $a_1$ in the sparse set $\mathcal{A}_\epsilon^{(1)}$. Then, $a_2$ and $a_3$ are excluded as stated in line 5 (center). Finally, $a_4$ and $a_5$ are added to $\mathcal{A}_\epsilon^{(1)}$ (right).

### 5.2 Revenue loss due to the sparse ad set approximation

In the following, we consider that the sparse approximation $\mathcal{A}_\epsilon^{(M)}$ is given as input to Volfied and show that in this case our algorithm will select a subset of ads that is almost as good as the one it selects when working with $\mathcal{A}$, with a much lower complexity. Specifically, our analysis consists of the following steps.
*(i)* We define the concept of *analogy* between two sets of ads, which expresses a similarity between the elements of the two sets with respect to both their position in the feature space, $\mathbb{F}$, and their value.
*(ii)* We show that, given a conflict-free set in $\mathcal{A}$, there always exists an analogue set in $\mathcal{A}_\epsilon^{(M)}$.
*(iii)* By starting from a slightly sparser subset of $\mathcal{A}$, we prove that there exists an analogue set in $\mathcal{A}_\epsilon^{(M)}$, which is also conflict free.
*(iv)* By exploiting this result, we quantify the maximum loss of revenue that there may be when $\mathcal{A}_\epsilon^{(M)}$ is used instead of $\mathcal{A}$. For each selected ad, such loss is equal to the contribution of the users located within a ring of radius $D_{\max}$ and width $\epsilon$ in the $\mathbb{F}$-space. Our results and the corresponding proofs are reported below; for ease of presentation, we refer to the case of a single PoA.

We first define the relationship of *analogy* between sets of ads extracted from $\mathcal{A}$ and $\mathcal{A}_\epsilon^{(M)}$, respectively.

***Definition 5.5.*** Set $\hat{\mathcal{S}}$ is analogue to set $\mathcal{S}$ ($\hat{\mathcal{S}} \approx \mathcal{S}$), if there exists a bijection function such that the image of every $a_i \in \mathcal{S}$ is an ad, $a_j \in \hat{\mathcal{S}}$, for which $D(a_i, a_j) \leq \epsilon$ and $r(a_j) \geq r(a_i)$.

Next, we show Step *(ii)*. The following theorem considers the case $M = 1$ and establishes that, for any conflict-free set in $\mathcal{A}$, there exists an analogue set that can be extracted from $\mathcal{A}_\epsilon^{(1)}$.

**Theorem 7.** Assume $M = 1$ and consider a set of ads, $\mathcal{S} \subseteq \mathcal{A}$, that is conflict free. If $D_{\max} > 2\epsilon$, then there exists a set $\hat{\mathcal{S}} \subseteq \mathcal{A}_\epsilon^{(1)}$ s.t. $\hat{\mathcal{S}} \approx \mathcal{S}$.

*Proof:* Because $\mathcal{S}$ conflict-free, for any arbitrary vehicle $v$, $\mathcal{S}$ includes at most a single ad, $a_i \in \mathcal{A}$, which is relevant to $v$, i.e., s.t. $D(a, v) < D_{\max}$. In order to build a set $\hat{\mathcal{S}} \subseteq \mathcal{A}_\epsilon^{(1)}$, which is analogue to $\mathcal{S}$, we proceed as follows.

Consider the ads in $\mathcal{S}$ one at a time, and denote the generic ad by $a_i$. If $a_i$ belongs also to $\mathcal{A}_\epsilon^{(1)}$, then the ad can be mapped onto itself. Otherwise, by construction, there exists some other ad, $a_j \in \mathcal{A}_\epsilon^{(1)}$, s.t. $D(a_i, a_j) \leq \epsilon$ and $r(a_j) \geq r(a_i)$. Hence $a_i$ can be mapped into $a_j$. Note that no other ad in $\mathcal{A}$ could have been already mapped onto $a_j$ because, in this case, there would exist a different ad, $a_q \in \mathcal{S}$, s.t. $D(a_q, a_j) \leq \epsilon$. However, since $D(a_i, a_j) \leq \epsilon$, due to triangle inequality, we would have: $D(a_i, a_q) \leq D(a_i, a_j) + D(a_j, a_q) \leq 2\epsilon \leq D_{\max}$. This contradicts the assumption on $\mathcal{S}$ being conflict free (recall that $M = 1$). Thus, we can obtain a set $\hat{\mathcal{S}}$ that meets the conditions stated by the theorem. □

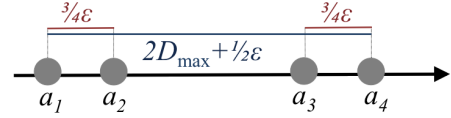

Fig. 3. Example with $n = 1$ (i.e., $\mathbb{F} \subseteq \mathbb{R}$) and $M = 1$: $\mathcal{S} = \{a_1, a_4\}$ is conflict free since $D(a_1, a_4) > 2D_{\max}$ (i.e., no user can find both $a_1$ and $a_4$ relevant to herself). Instead, $\hat{\mathcal{S}} = \{a_1, a_3\}$ is not conflict free as $D(a_1, a_3) = 2D_{\max} - \epsilon/4$.

The corollary below extends Theorem 7 to the case of denser sets of ads, $\mathcal{A}_\epsilon^{(M)}$ with $M > 1$, where each user may find up to $M$ ads relevant to herself.

**Corollary 1.** Consider $M > 1$ and a conflict-free set $\mathcal{S} \subseteq \mathcal{A}$. If $D_{\max} > 2\epsilon$, then there exists $\hat{\mathcal{S}} \subseteq \mathcal{A}_\epsilon^{(M)}$ s.t. $\hat{\mathcal{S}} \approx \mathcal{S}$.

*Proof:* The corollary is proved by induction. Initially, consider $M = 1$; by virtue of Theorem 7, the assertion holds. Then assume that the corollary holds for $m > 1$ and let us prove it for $M = m + 1$. To this end, recall that $\mathcal{A}_\epsilon^{(m+1)}$ is given by $\mathcal{A}_\epsilon^{(m)} \cup \texttt{EpsilonSet}(\mathcal{A} \setminus \mathcal{A}_\epsilon^{(m)}, \epsilon)$. Thus, $\mathcal{A}_\epsilon^{(m+1)}$ can be split into two disjoint sets $\mathcal{A}_\epsilon^{(m)}$ and $\mathcal{B}_\epsilon$, where $\mathcal{B}_\epsilon$ is obtained as sparse set of $\mathcal{B} = \mathcal{A} \setminus \mathcal{A}_\epsilon^{(m)}$. Similarly, $\mathcal{S}_{m+1}$ can be split into two disjoint sets. The first, $\mathcal{S}_m$, satisfies the induction hypotheses for $M = m$, and the second, $\mathcal{S}_1$, for $M = 1$. This means that there exist two disjoint sets in $\mathcal{A}_\epsilon^{(m+1)}$ that are analogue to $\mathcal{S}_m$ and $\mathcal{S}_1$, respectively. The union of such two sets is the set analogue to $\mathcal{S}_{m+1}$, which proves the corollary. □

The above result does not ensure that $\hat{\mathcal{S}}$ is conflict free. Indeed, as shown by the example in Fig. 3, the location in the $\mathbb{F}$-space of an ad in $\hat{\mathcal{S}}$ may vary with respect to its corresponding ad in $\mathcal{S}$ by as much as $\epsilon$, thus conflict freeness cannot be guaranteed in general. It follows that, although any conflict-free set $\mathcal{S} \subseteq \mathcal{A}$ has an analogue set $\hat{\mathcal{S}} \subseteq \mathcal{A}_\epsilon^{(M)}$, Volfied may never pick $\hat{\mathcal{S}}$ (as it is not conflict free). Only under some additional conditions, it is possible to show that $\mathcal{S} \subseteq \mathcal{A}$ does have its conflict-free analogue in $\mathcal{A}_\epsilon^{(M)}$.

Consider as maximum distance a slightly larger value than $D_{\max}$, i.e., $\Delta_{\max} = D_{\max} + \epsilon$, and let us define the concept of extended conflict freeness as follows.

***Definition 5.6.*** $\mathcal{S} \subseteq \mathcal{A}$ is an *extended conflict-free* set if, for every possible vehicle $v$, it does not include more than $M$ ads within $\Delta_{\max}$ from $v$.

Then we can prove the following important result: for any given extended conflict-free set, $\mathcal{S} \subseteq \mathcal{A}$, there exists an analogue set $\hat{\mathcal{S}} \subseteq \mathcal{A}_\epsilon^{(M)}$ that is also conflict free.

**Corollary 2.** Consider an extended conflict-free set $\mathcal{S} \subseteq \mathcal{A}$ and assume that $D_{\max} > 2\epsilon$. Then there exists a conflict-free set $\hat{\mathcal{S}} \subseteq \mathcal{A}_\epsilon^{(M)}$ such that $\mathcal{S} \approx \hat{\mathcal{S}}$.

*Proof:* By applying Corollary 1, we know that, given $\mathcal{S}$, there exists $\hat{\mathcal{S}}$ s.t. $\mathcal{S} \approx \hat{\mathcal{S}}$. What is left to show is that $\hat{\mathcal{S}}$ is conflict free. By contradiction, assume that $\hat{\mathcal{S}}$ is not conflict free. Then there exists a vehicle $v$ for which more than $M$ relevant ads are included in $\hat{\mathcal{S}}$. Since $\mathcal{S}$ is extended conflict free, it includes at most $M$ ads within distance $\Delta_{\max}$ from $v$. Then there must be an ad $a_i \in \mathcal{S}$, which is farther than $\Delta_{\max}$ from $v$, but its corresponding ad $a_j \in \hat{\mathcal{S}}$ is within $D_{\max}$ from $v$. Now, by assumption, $D(v, a_i) > \Delta_{\max} = D_{\max} + \epsilon$, and, since $a_i$ is mapped onto $a_j$, then $D(a_i, a_j) \leq \epsilon$. By triangle inequality, it is easy to see that this is in contradiction with the fact that $D(v, a_j) \leq D_{\max}$, which proves the assertion. □

In conclusion, $\mathcal{A}_\epsilon^{(M)}$ can provide an analogue conflict free set for any extended conflict-free set $\mathcal{S} \subseteq \mathcal{A}$. However, since extended conflict-free sets are a bit sparser than conflict-free sets, we may still have a performance loss, which increases with the value of $\epsilon$.

In order to quantify such loss we analyze the relation between the revenue obtained by selecting $\mathcal{S}$, denoted by $R(\mathcal{S})$, and that obtained with $\hat{\mathcal{S}}$, denoted by $R(\hat{\mathcal{S}})$. By virtue of Corollary 2, $\hat{\mathcal{S}}$ has very similar properties to $\mathcal{S}$. The only case where $R(\mathcal{S}) > R(\hat{\mathcal{S}})$ is when, given a vehicle $v$ and an ad $a_i \in \mathcal{S}$, $a_i$ is relevant to $v$, while its corresponding ad in $\hat{\mathcal{S}}$ is not. This case may occur when $a_i$ is almost at distance $D_{\max}$ from the vehicle in the $\mathbb{F}$-space. This motivates the definition below, which describes a conservative estimation of the obtained revenue that is preserved when considering $\mathcal{S}$ instead of $\hat{\mathcal{S}}$.

***Definition 5.7 (Conservative revenue estimation).*** Denote with $R^-(\mathcal{S})$ the conservative revenue estimation, which is the value of revenue computed by considering as relevant to a generic vehicle only those ads within distance $D_{\max} - \epsilon$ from the vehicle in the $\mathbb{F}$-space.

**Theorem 8.** *If $\mathcal{S} \subseteq \mathcal{A}$ is an extended conflict-free set, then there exists a conflict-free set $\hat{\mathcal{S}} \subseteq \mathcal{A}_\epsilon^{(M)}$ s.t. $R^-(\mathcal{S}) \leq R(\hat{\mathcal{S}})$.*

*Proof:* Consider all vehicles that contribute to $R^-(\mathcal{S})$. These vehicles are at most $D_{\max} - \epsilon$ far from their relevant ads in the $\mathbb{F}$-space, and, thus, in $\hat{\mathcal{S}}$ they are at most at distance $D_{\max} - \epsilon + \epsilon = D_{\max}$. It follows that they also contribute to $R(\hat{\mathcal{S}})$. Furthermore, every ad in $\hat{\mathcal{S}}$ has a value $r$ higher than or equal to that of its corresponding ad in $\mathcal{S}$, due to Corollary 2. Thus, any vehicle that contributes to $R^-(\mathcal{S})$, contributes a greater or equal amount to $R(\hat{\mathcal{S}})$. □

In conclusion, when Volfied is applied to $\mathcal{A}_\epsilon^{(M)}$ instead of $\mathcal{A}$, we can guarantee that it has the opportunity to select a set of ads, which, in the worst case, is almost as good as that it would select by using $\mathcal{A}$. Indeed, for each selected ad, the revenue loss is limited to the contribution of the users located within a ring of radius $D_{\max}$ and width $\epsilon$ in the $\mathbb{F}$-space. In the cases where, using $\mathcal{A}$, Volfied generates an extended conflict-free set of ads, then there is no revenue loss in using $\mathcal{A}_\epsilon^{(M)}$.

# 6 PERFORMANCE EVALUATION

We first describe the real-world reference scenario we consider to assess the performance of Volfied and to compare it against two benchmark strategies (Sec. 6.1). The performance results are then presented when ads are not cached (Sec. 6.2), as well as when vehicles can cache ads and display them at a later time (Sec. 6.3). Interestingly, our results show that Volfied outperforms Top-k in both scenarios.

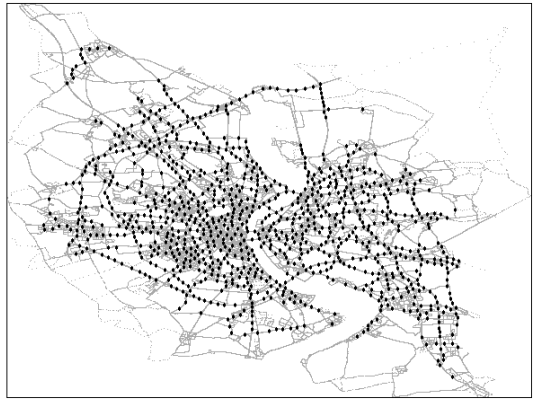

Fig. 4. Road layout (gray lines); deployed RSUs are represented by black dots.

## 6.1 Reference scenario

We evaluate Volfied using a vehicular trace [48] depicting car mobility in Cologne, Germany. The trace refers to a urban area of $28 \times 32$ km$^2$, and models over 110,000 vehicles, during the course of 8 hours. We consider RSUs as PoAs and place 1,000 RSUs along the busiest roads and at the center of intersections, as shown in Fig. 4. More sophisticate RSU deployment techniques, e.g., [35], [39], [40], [41], [42], [43] discussed in Sec. 2.3, could be adopted as well, requiring no change to our decision-making schemes. The effect of a sparser RSU deployment, namely, when only 100 RSUs are placed over the reference area, is instead considered in Sec. 6.3.

The RSU range is set to 150 m, which, in the default case of 1,000 RSUs, results in covering roughly half of our reference area. We stress that such a deployment is challenging for Volfied, since it leads to a scenario where vehicles are under radio coverage for just about 60% of the time they appear in the trace.

Each vehicle is assigned a five-dimensional feature vector, i.e., $n = 5$. Feature values are sampled from the normal distribution with a mean of 0.5 and standard deviation of 0.15. Similarly, each ad is assigned a five-dimensional feature vector and a value $r$, both sampled uniformly in the range $(0, 1)$. 90% of the ads are global, the others are local. Recall that local ads can be displayed only within the coverage area of a specific RSU, which is selected at random.

We evaluate Volfied, Top-k and a Random strategy by simulating the system over 480 time steps, with each step lasting one minute. At each time step, RSUs broadcast the ads selected by the broker. The tested algorithms only differ from each other in the ad selection strategy, i.e., all of them can access the same information on ads, vehicles and ad values. Specifically, Top-k selects the $K$ ads with highest estimated revenue $R$, Random selects $K$ random ads among those that have a positive revenue $R$, and Volfied makes conflict free selection as described in Sec. 4.1.

We first assume in Sec. 6.2 that vehicles do not cache ads, i.e., $C = 0$. Thus, in general, whenever a vehicle receives more than $M$ ads, i.e., more ads than can be displayed in a time step, it displays the $M$ most relevant ones and discards the others. Notice that this never happens with Volfied, which makes conflict-free decisions. In Sec. 6.3, instead, we consider that, whenever a vehicle receives more than $M$ relevant ads, it displays the $M$ most relevant ones and caches the others (up to $C$, in order of decreasing relevance). The vehicle may then display the cached ads in future time steps. Note that the content of the cache is decided at each time step individually, accounting for both the newly received ads and the ones in the cache. Also, cached ads that are no longer relevant, due to the changed vehicle location, are dropped, and so are the ads that have been already displayed by the vehicle.

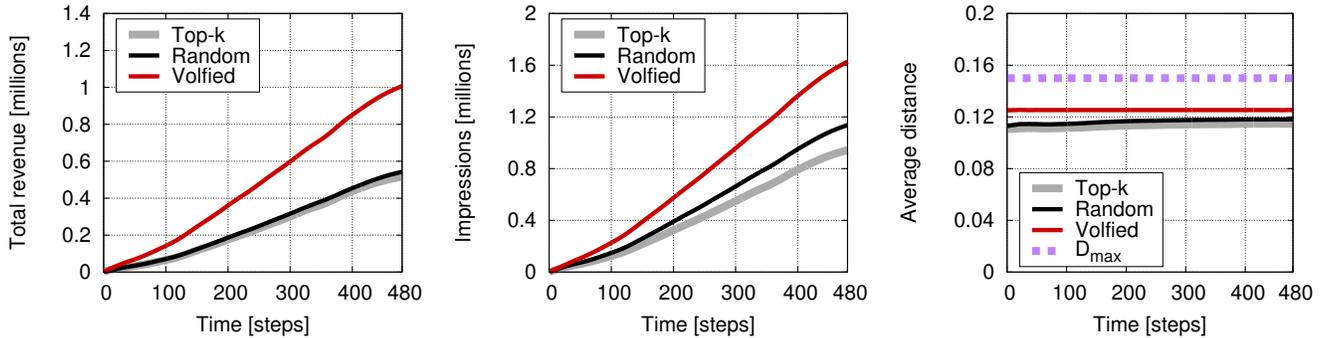

Fig. 5. Time evolution of the cumulative revenue, cumulative number of impressions and average distance for Volfied, Top-k and Random (default configuration).

With regard to the system parameters, most of the results are derived under our *default configuration*, whose settings are: $K = 5$, $M = 1$, $|\mathcal{A}| = 10,000$, $D_{\max} = 0.15$, and $\epsilon = 0.025$, and under the assumption that the presence of vehicles under an RSU can be detected without error. Note that the default configuration implies that the tested algorithms use as input the sparse set $\mathcal{A}_{0.025}^{(1)}$. Also, unless stated otherwise, the Euclidean distance (1) is used as definition of $D(\vec{f}_1, \vec{f}_2)$.

The performance metrics we plot are:
- total revenue, which reflects the amount of money paid to the broker by advertisers and is computed as the sum of the revenue generated by all broadcasted ads (recall that the revenue is equal to the ad value $r$ multiplied by the number of ad impressions);
- total number of impressions, i.e., the total number of ads that have been displayed by vehicles. This metric reflects the point of view of advertisers who would like to maximize ad visibility;
- the average impression distance, which represents how relevant, on average, a displayed ad is to the user. This last metric accounts for the user's point of view. The lower the distance, the more relevant the displayed ads are to the users. The average distance never exceeds $D_{\max}$.

We remark that other performance metrics such as bandwidth consumption are virtually the same for all tested algorithms and are therefore omitted.

TABLE 2
Comparison against the optimum for a single time step

| Metric | Top-k | Random | Volfied | Optimum |
|---|---|---|---|---|
| Revenue | 1444.3 | 810.1 | 1712.0 | 1770.3 |
| Impressions | 1647 | 1573 | 1910 | 1889 |
| Distance | 0.107 | 0.115 | 0.125 | 0.119 |

### 6.2 Numerical results with $C = 0$

**Comparison against the optimal solution.** We first compare the performance of Volfied, as well as that of Top-k and Random, against the optimum derived through (3)–(8). To this end, we restrict ourselves to a single-step scenario so that the computation of the optimal solution is viable. The results in Tab. 2 show that Volfied provides near-optimal performance: its revenue is just $3.4\%$ lower than the optimum, while it generates $1\%$ more impressions and similar distance. Note that, since the optimum maximizes the revenue, there may be cases where it selects ads with very high value $r$ but that are displayed by slightly fewer users, while Volfied always generates a conflict-free set thus resulting in a higher number of impressions. The performance gap between the optimum and the other two schemes is much larger: the revenue gain is 25% and 55% when compared to Top-k and Random, respectively.

**Performance over time.** Fig. 5 shows the time evolution of our performance metrics for the default configuration. As can be observed, Volfied generates 70% higher revenue and 50% more impressions than Top-k. This implies that Volfied satisfies the interests of both broker and advertisers. Because Volfied aims at maximizing the broker's revenue, it may select ads that are slightly (by about 0.01) less relevant to users with respect to Top-k and Random, as shown by the right plot in Fig. 5.

**Effect of the ad set size.** Fig. 6 shows the impact of the ad set size, $|\mathcal{A}|$, on the system performance. Intuitively, the larger $|\mathcal{A}|$, the easier it is to find relevant ads to each vehicle. Indeed, revenue (and also impressions, omitted for brevity) improves for larger values of $|\mathcal{A}|$. Interestingly, for 1,000 ads the difference between the algorithms is small and Volfied generates 30% more revenue than its alternatives. However, as $|\mathcal{A}|$ increases, the performance gap also grows, and when $|\mathcal{A}| = 20,000$, Volfied increases revenue by 70% with respect to the other schemes. Indeed, the more ads in the system, the more critical their selection becomes and the more severe the revenue loss that occurs due to the conflict discussed in Sec. 3.3. Hence the advantage provided by Volfied becomes more evident.

**Number of ads to serve.** The value of $K$ corresponds to the bandwidth that is consumed by ad broadcasting. The left plot in Fig. 7 shows that, for small values of $K$, the broker's revenue increases with $K$. However, it is interesting to notice that the revenue saturates as eventually the vehicles' ability to display ads and the number of relevant ads become a performance bottleneck. Thus, there is a preferred value of $K$ (which depends on the system settings) that the broker should use.

Furthermore, it is surprising to notice that the performance of Top-k and Random is not monotone with $K$: increasing $K$ beyond a certain point actually hurts the broker's revenue. The reason is twofold. First, the larger $K$, the more likely the conflicts. Second, ads that were broadcasted before are not considered as profitable anymore (although they can still generate revenue if not all vehicles displayed them); thus, once the top $K$ ads have been broadcasted, it becomes increasingly harder to identify the best ads to transmit. Interestingly, Top-k reaches its peak value of revenue for a lower $K$ than Random, due to the fact that the ads selected by Top-k create conflicts more often than those that are randomly chosen. This is confirmed by the right plot in Fig. 7, which shows that the conflicts generated by Top-k reduce the distance between ad and user, hence providing slightly smaller average distance than Random.

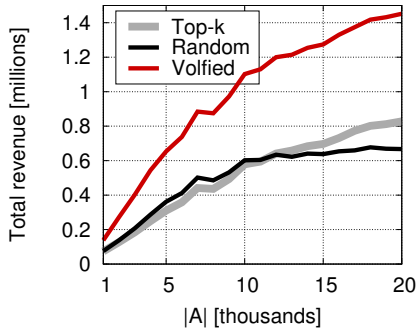
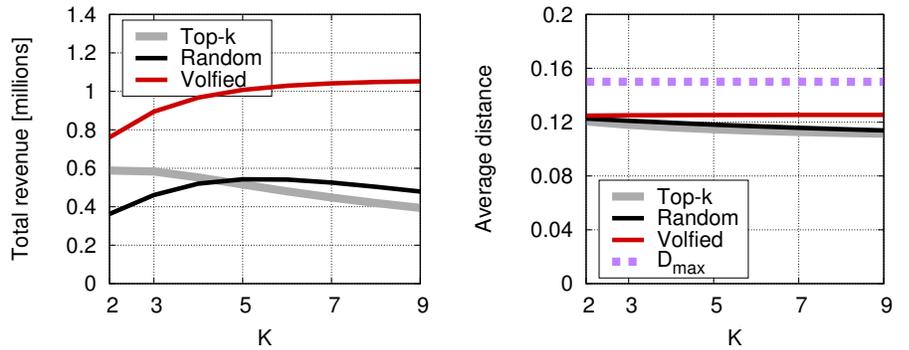

Fig. 6. Effect of the ad set size on broker's revenue, when $K = 5$ and $M = 1$.

Fig. 7. Effect of number of broadcasted ads ($K$) per time step, with fixed $M = 1$.

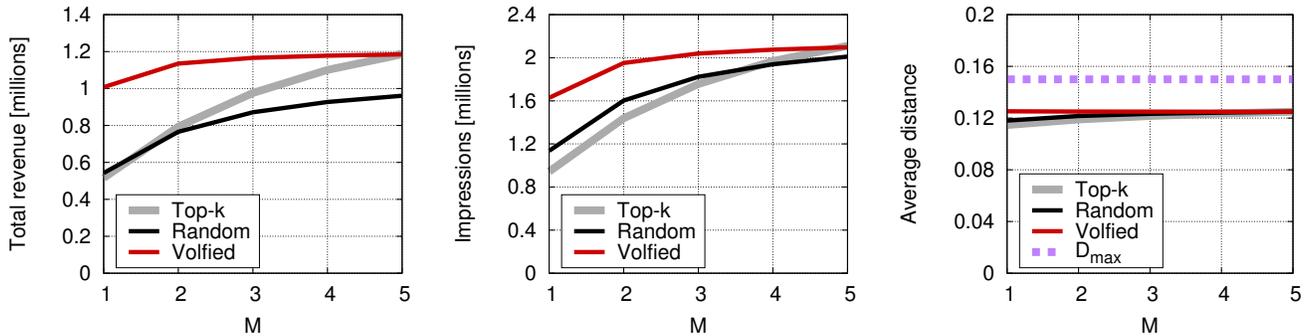

Fig. 8. Effect of the number of displayable ads ($M$) on performance metrics for fixed $K = 5$.

**Effect of $M$.** We now fix $K = 5$ and study the performance as $M$ varies. Recall that a small value of $M$ accounts for the reduced screen size aboard a vehicle and for the limited driver's attention span, and that typically $M < K$. As shown in Fig. 8, Top-k and Random are very sensitive to $M$. For $M < K$, they provide much lower revenue and number of impressions; only when $M$ approaches $K$, i.e., when conflicts seldom occur, Top-k gives good performance. Volfied, instead, is much more robust, as its performance varies very little with $M$. It generates just $10\%$ lower revenue and $15\%$ fewer impressions when $M = 1$ than when $M = K$. We remark that the latter is the special case where Volfied and Top-k yield the same revenue, which coincides with the optimum (see also Sec. 4.1). Thus, the fact that Volfied gives a similar revenue for $M = 1$ and for $M = K$, confirms that its performance is near-optimal for any $M < K$.

**Effect of $\epsilon$.** The top plot in Fig. 9 depicts the broker's revenue as $\epsilon$ varies, when $D_{\max} = 0.15$. As can be observed, for values of $\epsilon \leq D_{\max}/4$, the revenue loss due to the sparse approximation is negligible. Also, for such values of $\epsilon$ we limit the number of ads that have to be processed per vehicle arrival/departure. As shown in Theorem 6, the number of processed ads is bounded by: $\left\lceil \left(\frac{D_{\max}}{\epsilon}\right)^n \right\rceil$ regardless of the size of the ad set $\mathcal{A}$, which, for $\epsilon = D_{\max}/4$, amounts to $4^5 = 1024$.

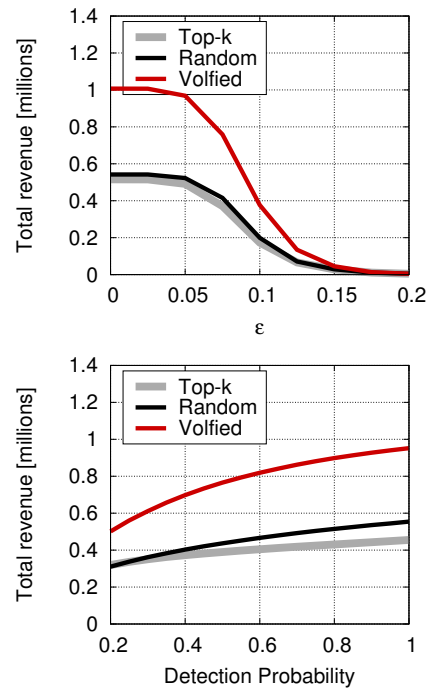

Fig. 9. Effect of $\epsilon$ (top) and of accuracy in vehicle detection (bottom) on the broker's revenue.

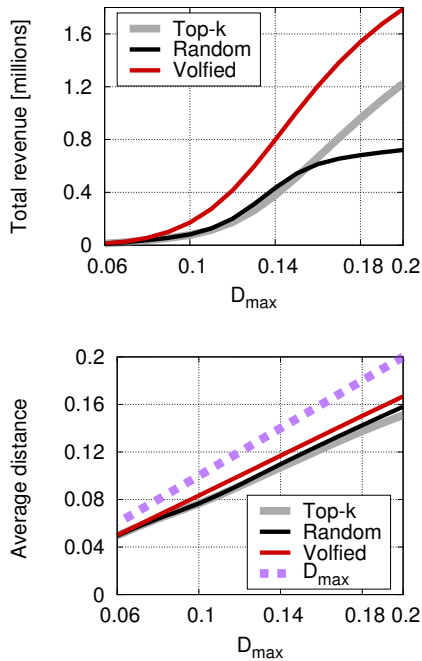

Fig. 10. Effect of $D_{\max}$ on performance metrics.

**Vehicle detection accuracy.** While deriving the previous results, we assumed that an RSU could reliably detect all vehicles under its coverage thanks to their beacon messages. The bottom plot in Fig. 9 shows the impact of different levels of accuracy, i.e., probability of successfully detecting a vehicle under an RSU. Remarkably, Volfied with 0.3 accuracy provides higher revenue than Top-k with accuracy equal to 1. It follows that Volfied is very effective even with incomplete knowledge of the scenario, since it can still successfully avoid conflicts.

$D_{\max}$ **and average distance.** $D_{\max}$ is another important parameter as it determines which ads are relevant to a user. Intuitively, the larger $D_{\max}$, the easier it is to select relevant ads that will be displayed by a vehicle, but also the larger the average ad-user distance. Fig. 10 confirms these trends for all selection strategies. However, we can see that, when $D_{\max}$ is very small, all strategies yield similar revenue and average distance as the set of ads with positive revenue, hence that can be selected, is very small. Likely each vehicle has at most one ad within distance $D_{\max}$ in $\mathcal{A}_\epsilon^{(1)}$. For larger values of $D_{\max}$, instead, Volfied provides higher revenue than the other schemes, as conflicts become increasingly likely and cause revenue loss (top plot in Fig. 10). For Volfied, the price to pay is a slight increase in the average ad-user distance (bottom plot in Fig. 10).

**Impact of the distance metric.** So far, we have used the Euclidean distance defined in (1) as the definition of $D(\vec{f_1}, \vec{f_2})$. In Fig. 11 we check if, and to which extent, using the angular distance defined in (2) changes the relative performance of Volfied and its counterparts. Notice that, as the definition of distance changed, we need to use a different threshold; specifically, we set $D_{\max} = $ 0.092 as this guarantees the same average number of relevant ads per vehicle as in the default scenario.

Fig. 11 (left) and Fig. 11 (center) are the counterparts of Fig. 5 (left) and Fig. 5 (center), respectively, and depict the time evolution of the revenue and number of impressions. We can see that the relative performance of Volfied and its counterparts is unchanged: Volfied guarantees both a higher revenue to the broker and a higher number of impressions compared to Top-k and Random. Fig. 11 (right) shows how the number $M$ of ads that can be displayed impacts the broker revenue, when $K = 5$, and is the counterpart of Fig. 8 (left). We can again see that changing the distance metric does not substantially alter the qualitative and quantitative behavior of the schemes under study.

### 6.3 Numerical results with $C > 0$

We now consider that vehicles can store ads in their cache and display them at a later time if convenient. Note that cache availability and size impact the performance of Volfied's counterparts only, most notably Top-k. Indeed, Volfied always delivers a set of ads that can be immediately displayed, i.e., it does not exploit caches even if they are available. Top-k instead would become the optimal strategy when the vehicles are equipped with unlimited cache and all the ads are global, as under these conditions every transmitted ad will eventually be displayed – with caching having the high-level effect of making up for the conflicts generated by Top-k. In practice, however, caches are bounded in size and local ads can still be dropped from the cache if a vehicle leaves their relevance area. Therefore, Volfied retains its performance advantage, as shown below.

**Effect of cache size $C$.** Fig. 12 depicts the total revenue and the number of impressions as the vehicles cache size varies. The latter is normalized to the average number of existing ads that are relevant to a generic vehicle when $\mathcal{A}_{0.025}^{(1)}$ is considered as ad set (namely, 5 ads). It is clear that the performance of both Random and Top-k dramatically improves since the caching mechanism mitigates the problem of conflicting ads. Indeed, at a given time step, a vehicle can store the ads that are not relevant enough to be displayed immediately, and display them later when their relevance outweighs that of other newly received ads, or the vehicle happens to be out of RSU coverage.

**Sparse deployment.** In order to assess the performance of the different solutions in a scenario where caching plays a major role, in Fig. 13 we focus on an extremely sparse scenario, including 100 RSUs only. This turns into vehicles being under RSU radio coverage just for about 7% of the time they appear in the trace, thus leaving plenty of opportunities for the vehicles to display cached ads. Consistently, the number of impressions achieved by Top-k (see bottom plot) gets much closer to the one of Volfied, with respect to the case shown in Fig. 12; for large values of cache size, it even exceeds the number of impressions obtained under Volfied. However, looking at the total revenue (top plot), we observe that, in this extreme scenario, Volfied and Top-k tend to provide the same performance. We also wish to recall that in the default configuration 90% the ads are global and therefore cacheable – another condition that favors Top-k over Volfied. Finally, comparing Fig. 13 to Fig. 12, it is worth noticing that the performance of all schemes significantly decreases in the presence of such a sparse RSU deployment.

**Effect of $|\mathcal{A}|$ and $K$.** We now fix the normalized cache size to 1.4 and investigate the impact of the ads set size (Fig. 14) and of the number $K$ of ads broadcasted by each RSU (Fig. 15). In such a scenario, Top-k and Random show an improvement but Volfied still greatly outperforms both of them confirming that caching cannot fully solve the problem of ad conflicts. Focusing on Top-k and Random, we also observe that in general the performance of Random is either the same or lower than Top-k. Indeed, when RSUs have chance to transmit many ads, Random only selects the ones with positive revenue, thus ending up choosing the same ads as Top-k. When instead there is no room for all ads with positive revenue, then using Top-k makes a difference with respect to the random choice. Furthermore, it is interesting to note that, as shown in Fig. 15, the performance of both Top-k and Random decreases as $K$ grows beyond a certain value. The reason is that, as RSUs broadcast more ads than what users can display or cache, some ads are dropped by the vehicles. However, since no feedback is provided, RSUs are unaware of which ads get discarded. This implies that RSUs do not rebroadcast ads that it could be convenient to deliver again to the vehicles, leading to a severely suboptimal ad selection.

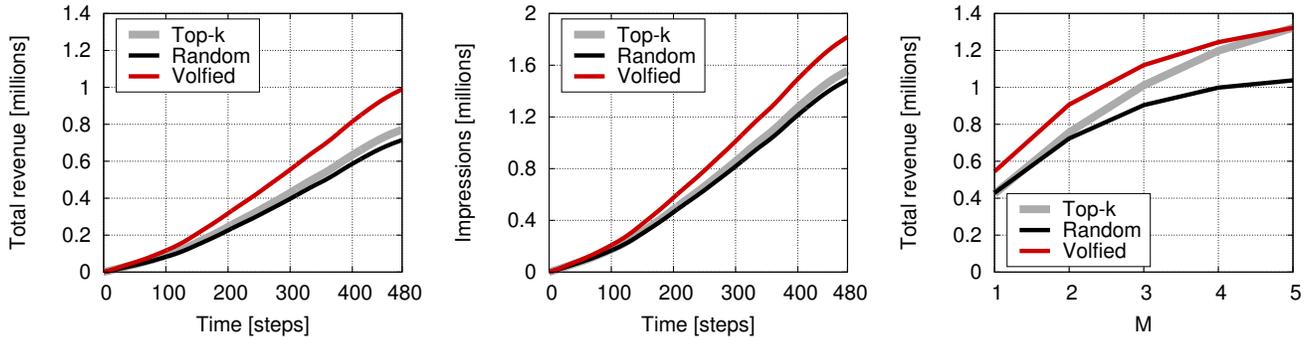

Fig. 11. Effect of using the angular distance (2) as the definition of $D(\vec{f_1},\vec{f_2})$: time evolution of the revenue (left) and number of impression (center), and effect of $M$ on the revenue when $K=5$ (right).

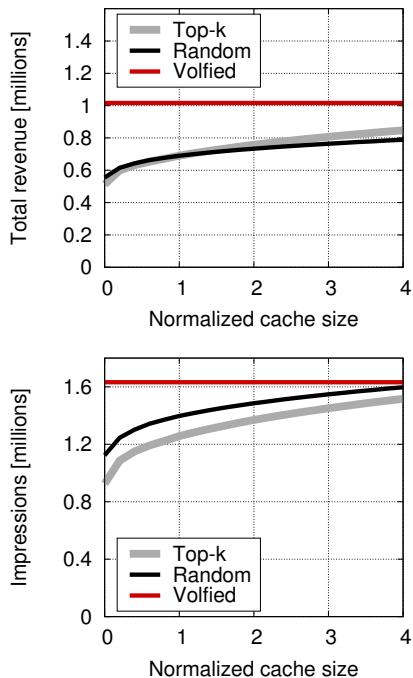

Fig. 12. Effect of the cache size on the performance metrics, under the default configuration. The cache size is normalized to the average number of per-user relevant ads.

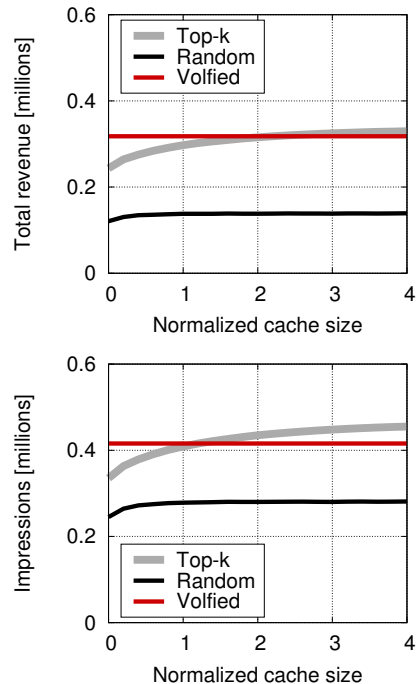

Fig. 13. Total revenue and number of impressions in the presence of a sparse RSU deployment, where vehicles spend under RSU radio coverage just 7% of the time they appear in the trace. Default values are assumed for the other system parameters.

## 7 CONCLUSIONS

We addressed targeted advertising in vehicular networks and envisioned a system where advertisers pay a broker based on the value and the number of impressions of each ad. We considered the broker's perspective and formulated the problem of selecting the ads to broadcast that maximize the broker's revenue, subject to a maximum number of ads that can be transmitted. While doing this, we identified a conflict between user and broker's interests, which severely hurts the broker's revenue if not properly addressed. Then, in light of the problem complexity, we introduced Volfied, an efficient greedy algorithm that always selects a conflict-free set of ads while maximizing the broker's revenue. The complexity of Volfied has been proved to be linear with the number of ads.

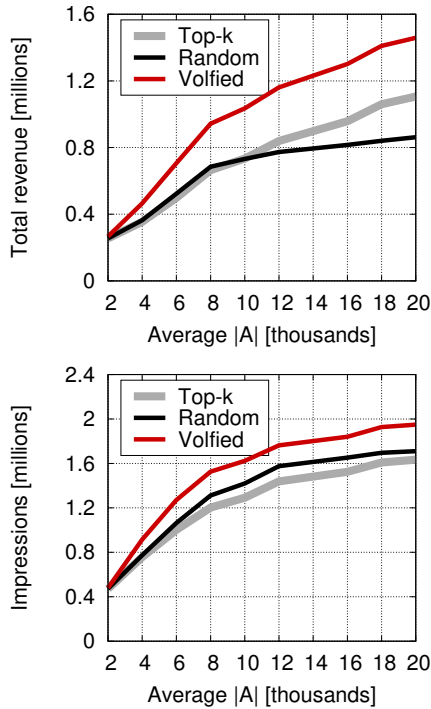

Fig. 14. Effect of varying the ad set size, on the broker's revenue and the number of impressions ($K=5$, $M=1$, and normalized cache size equal to 1.4).

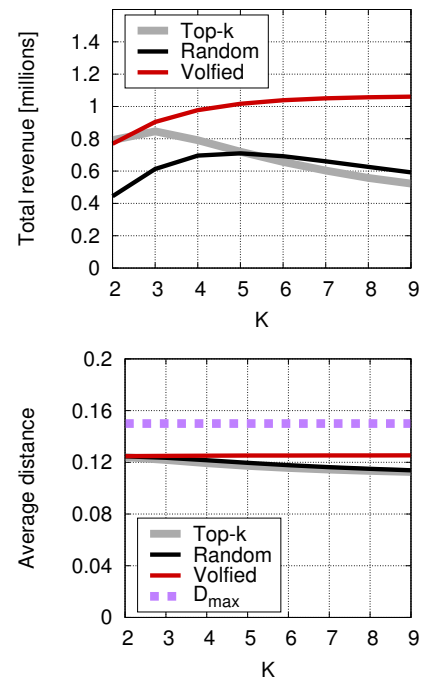

Fig. 15. Effect of number of broadcasted ads ($K$) per time step with $M=1$ and normalized cache size equal to 1.4.

In addition, we proposed a sparse approximation of the ad set, which further speeds up ad selection. We evaluated Volfied and our sparse approximation technique in a realistic vehicular environment, against the optimum in a single-time step scenario and against the Top-k and Random strategies in a multi-time step scenario. Numerical results show that Volfied provides near-optimal performance. Also, it improves the broker's revenue by up to 70%, and the number of displayed ads by up to 50%, with respect to Top-k.

Finally, we extended our analysis to the case where vehicles are equipped with a cache. Our results show that, even if Volfied does not exploit the vehicles' caching capabilities, its performance still matches or outperforms the one of Top-k. This further highlights how Volfied represents an effective and efficient solution for vehicular advertising, with low hardware requirements on vehicles and low implementation complexity.

## ACKNOWLEDGEMENT

This work is supported by the European Commission through the H2020 5G-TRANSFORMER project (Project ID 761536).

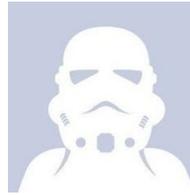

**Gil Einziger** is.

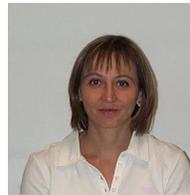

**Carla-Fabiana Chiasserini** (M'98, SM'09, F'18) graduated in Electrical Engineering (summa cum laude) from the University of Florence in 1996. She received her Ph.D. from Politecnico di Torino, Italy, in 2000. She has worked as a visiting researcher at UCSD in 1998–2003, and she is currently an Associate Professor with the Department of Electronic Engineering and Telecommunications at Politecnico di Torino. Her research interests include architectures, protocols, and performance analysis of wireless networks. Dr. Chiasserini has published over 250 papers in prestigious journals and leading international conferences, and she serves as Associated Editor of several journals.

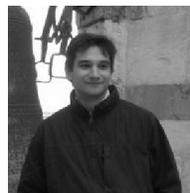

**Francesco Malandrino** earned his Ph.D. in 2012 from Politecnico di Torino, Italy, where he is currently a researcher. Before his current appointment, he held short-term positions at Trinity College, Dublin, and at the Hebrew University of Jerusalem as a Fibonacci Fellow. His interests focus on wireless and vehicular networks and infrastructure management.